# FoVR: Attention-based VR Streaming through Bandwidth-limited Wireless Networks

Songzhou Yang*, Yuan He†, Xiaolong Zheng‡
*†School of Software and BNRist, Tsinghua University
‡School of Computer Science, Beijing University of Posts and Telecommunications
*yangsz18@mails.tsinghua.edu.cn, †heyuan@tsinghua.edu.cn, ‡zhengxiaolong@bupt.edu.cn

*Abstract*—Consumer Virtual Reality (VR) has been widely used in various application areas, such as entertainment and medicine. In spite of the superb immersion experience, to enable high-quality VR on untethered mobile devices remains an extremely challenging task. The high bandwidth demands of VR streaming generally overburden a conventional wireless connection, which affects the user experience and in turn limits the usability of VR in practice. In this paper, we propose FoVR, attention-based hierarchical VR streaming through bandwidth-limited wireless networks. The design of FoVR stems from the insight that human's vision is hierarchical, so that different areas in the field of view (FoV) can be served with VR content of different qualities. By exploiting the gaze tracking capacity of the VR devices, FoVR is able to accurately predict the users attention so that the streaming of hierarchical VR can be appropriately scheduled. In this way, FoVR significantly reduces the bandwidth cost and computing cost while keeping high quality of user experience. We implement FoVR on a commercial VR device and evaluate its performance in various scenarios. The experiment results show that FoVR reduces the bandwidth cost by 88.9% and 76.2%, respectively compared to the original VR streaming and the state-of-the-art approach.

## I. INTRODUCTION

With the booming development of the Virtual Reality (VR) technology, consumer VRs [1]–[3] with attractive characteristics like immersion, interaction and imagination, are promising to open a new era of industry [4], entertainment, education and etc.. The global VR market is estimated to be $26.89 billion by 2020 [5]. According to a market report [6], more than 70% of Internet traffic is consumed by video streaming in the United States. In existing VR applications, 360° video contents occupy an absolute dominant position in current VR contents, accounting for up to 99.37%, according to the report from Huawei [7]. We can envision that the 360° video streaming is going to dominate Internet traffic in the near future.

In spite of the rapid growth of VR market, there are huge gaps between the limited capacity of existing infrastructure and the high demand of VR contents, especially for the high-end mobile VR streaming as known as 360° video streaming. First, a huge gap exists between the bandwidth capacity of conventional wireless technologies and the bandwidth demand of 360° video streaming while wireless networks are widely used in mobile applications [8]. According to the Pixels Per Inch (PPI) analysis $a = 2arctan(\frac{h}{2})$, where $a$ is the Field of View (FoV), $h$ is the pixel pitch and $d$ is the distance from eye to screen, if we want to achieve a retina display on VR devices, we need 5073×5707 resolution for one eye (estimated by simulating a screen with 20m distance to eyes with 95° FoV). Hence, the bandwidth demand is at least 840Mbps when we use the coding mode of 4K video and at least 4.2Gbps if we desire the 3D experience of the VR video with 120 Frame Per Second (FPS). However, the bandwidth of the fastest commercial wireless network, WiFi(802.11ac), is 1.3 Gbps in theory and can only achieve 400 Mbps data rate in practice [9]. Second, the limited hardware resource on VR devices cannot meet the high decoding demand of 360° video streaming with high quality. The existing solutions of high-end VR systems usually attach the high-end VR to a powerful PC which is connected to the Internet via wired connections, as called as tethered VRs. The PC also provides abundant computation resources to decode the high-quality streaming to avoid the high decoding delay on the VR devices. However, the wired connection restricts the user mobility and may even cause trouble or accidents, e.g. stumble. Streaming 360° videos via a conventional wireless connection can only obtain low quality of experience (QoE).

To fill those gaps and provide high-quality VR contents on mobile VR devices, recent works have made attempts in the following three ways: (1) adopt new wireless technologies like 60GHz mmWave [10]–[12] to enhance the network capacity; (2) adopt video compression techniques like tile-based streaming to reduce the size of VR streams [13]–[16]; (3) adopt pre-computing and offloading techniques to relieve local computing loads so that high-quality content can be obtained on the resource-limited VR devices [17]–[19].

However, these solutions still have shortcomings and cannot meet the requirement of 360° video streaming. The new wireless technologies using mmWave is sensitive to device mobility, which may suffer performance degradation when user moves. The emerging wireless network technologies such as 5G and WiGig can only relieve but not solve the problem. Tile-based video compression methods sacrifice the user's QoE to save bandwidth cost and cannot compress the high quality 360° videos to a satisfying size. Offloading-based schemes work for only heavy computing tasks like VR games where the desired content can be computed at the host or cloud and then transmitted to the client without bandwidth limitation. However, for 360° video streaming, the compressed video has to be decompressed at the mobile devices because of the limited bandwidth. In a nutshell, high-quality 360° video streaming on mobile VR devices through conventional wireless connections is still a challenging and urgent task.

In this paper, we propose FoVR, an attention-based hierarchical 360° video streaming system through bandwidth-limited wireless networks. FoVR leverages the hierarchical property of human vision, predicts the user's attention, and accordingly schedules the mixed-quality 360° video streaming to a VR HMD (Head Mounted Device). Specifically, FoVR divides a user's vision into three hierarchical areas, where the user's perception of visual information actually demands different levels of video quality. Hence, FoVR utilizes the head and gaze movement information provided by the HMD to detect and predict the user's FoV and attention. Based on the prediction, streaming of the mixed-quality VR content is scheduled so as to meet all the bandwidth constraints, computing resources requirement and QoE of high quality. The main contributions of this work can be summarized as follows.

- We propose FoVR, a hierarchical structure of 360° video streaming on mobile VR HMD. The design of FoVR exploits the humans hierarchical vision and composes mixed-quality VR clips, with a premise of saving bandwidth while maintaining a high QoE.
- We address the technical challenges. Specifically, we design a support vector regression based method for prediction of users attention. Takes advantages of accurate prediction, FoVR optimizes the scheduling of 360° video streaming, so that the limited wireless bandwidth is sufficiently utilized.
- We implement FoVR on commercial VR HMD and conventional WiFi networks. We extensively evaluate FoVR in many scenarios. The evaluation results demonstrate that FoVR reduces the bandwidth cost by 88.9% and 76.2% in average, respectively compared to the original 360° video streaming and the state-of-the-art approach.

The rest of this paper is organized as follows. Section II presents the related works. In Section III, we conduct some background knowledge and the preliminary studies. We introduce the design of FoVR in Section IV and evaluate the performance of FoVR in Section VI. We conclude our work in Section VII.

## II. Related Work

To enable the high-quality contents on mobile VR devices, recent works focus on increasing the bandwidth of wireless networks and reducing the video size to narrow the gap between the bandwidth capacity and the bandwidth demands. To avoid running complicated decoding algorithms on the resource-limited mobile VR devices, reducing the computation load is also studied.

**Bandwidth capacity improvement.** TPCast [10] and DisplayLink XR [11] are commercial VR add-ons that utilize the 60GHz mmWave to enlarge the network bandwidth. MoVR [12] propose a configurable mmWave mirror to amplify and reflect the mmWave signal to the receiver. Although these methods achieve the untethered VR, they are not easy to implement on mobile VRs due to the deployment cost of the infrastructure. Besides, these methods still cannot enable the direct access to server on the Internet. They only replace the HDMI cables and still need a nearby PC as the helper for decoding and transmissions.

**Bandwidth demands reduction.** Tile-based streaming tries to reduce the bandwidth demands. The authors in [13] propose to deliver the predicted visible portion of 360 videos to reduce the video size. The authors in [14] propose transmitting the tiles in FoV with higher priority to improve QoE. In [15], the authors uses spatial compression based on the user's region-of-interest (ROI) and rate control to reduce the video size. Based on the inferred user's FoV, tile-based streaming methods delivery video tiles only in FoV and discard the OoS (Out of Sight) tiles. However, due to the large area of FoV (120°for two eyes), the reduction is limited. Different from those methods, FoVR leverages the hierarchical property of human vision and provides video tiles with different definitions even in the FoV to greatly reduce the size of video. Besides, inferring user's attention by head movement implicitly assume the center of head orientation as the attention direction, which may not hold in practice because the eyes move more frequently than the head and do not necessarily look at the center. Hence, existing tile-based streaming methods also has unsatisfied QoE due to the frequent watching interrupts caused by inaccurate inference.

**Computing loads reduction.** Pre-computing and offloading technologies have been proposed to reduce computing loads of VR devices. Flashback [17] utilizes pre-computing and caching to eschews real-time scene rendering. Furion [18] and Cutting the Cord [19] leverage offloading to relieve local computing loads. Foveated rendering [20] also leverages the hierarchical human vision but focuses on reducing the computation cost of rendering. These works focus on the scenario where computation resource is limited but network bandwidth is sufficient. But 360° video streaming that our work focuses on is bandwidth-hungry. There is no surplus bandwidth for offloading. The first thing in 360° video streaming is heavy bandwidth demand. Reduction of bandwidth demands will bring reduction of computation demands naturally.

In summary, different from existing works, FoVR leverages the hierarchy of human vision to display contents with different definitions in different grids of FoV. Hence, we guarantee the integrity of videos to maintain a satisfied QoE. Also, FoVR continuously predicts the hierarchically visible field and schedules the composition and prefetching of the mixed-quality video, which adapts to dynamic network conditions.

## III. Background and Preliminary Study

In this section, we will introduce some background knowledge about 360° video streaming and human vision. We also present the preliminary analysis of our work.

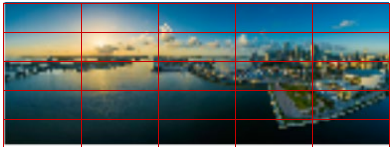 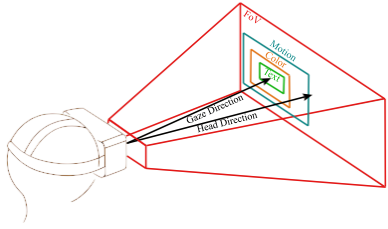 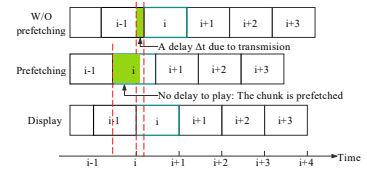

Fig. 1. A 360° frame with gridding.

Fig. 2. Human's vision in the VR environment.

Fig. 3. Streaming display delay with prefetching and without prefetching.

### A. VR Background

VR HMDs are classified into three types, tethered VR that are attached to a PC, mobile VRs that are attached to a phone, and standalone VRs that have all functions without a host. Using build-in sensors such as IMUs (Inertial Measurement Unit) and infrared ray (IR) sensors, most of the VRs provide orientation tracking and position tracking ability, which are essential for delivering video clips only in FoV. Furthermore, there are some VRs and VR add-ons provide eye tracking or gaze tracking, such as Tobii [21] and Fove VR [22]. We can also detect the gaze movement of the users.

360° videos are captured by a dedicated system in which multiple cameras simultaneously record on overlapping angles. The videos are then integrated into one 360° video clip by video stitching. There are two major formats of 360° videos: Equirectangular projection and Cubemap projection. In this paper, we consider the equirectangular projection, the most common format in practice. Figure III shows an example of an equirectangular 360° video frame with grids.

As the name suggests, a 360° video has 360° field of view in both horizontal and vertical level. But when we watch a 360° video, we can only see a fraction of the whole video at one moment, while the rest is invisible. In existing implementations, both the visible and invisible parts are encoded in the video streaming because we cannot know which parts are visible. But we should notice that the invisible part occupies bandwidth as well, which even consume much more bandwidth than the visible part, as analyzed in Section III-B.

### B. Human Vision

As shown in Figure III, human has a two-eye horizontal FoV up to 190°, of which about 120° makes up the Binocular FoV [23], which is the visible to both eyes. The vision is intrinsically hierarchical [24], mainly including central vision, paracentral vision, and peripheral vision, corresponding to text area, color area and motion area in Figure III, respectively. *Central vision*, also called as fovea vision, is the most important part for human vision. Central vision only has about 5° of human vision. In central vision, a person can clearly see in text focus with about four or five words with 100% accuracy. *Paracentral vision* has 30°, which is a bit larger than the central vision. In the field of paracentral vision, people can perceive the color of objects. *Peripheral vision*, occupying about 60° of human vision, gives a perception of motion.

Thanks to the hierarchical human vision, we notice the chance to display contents with different definitions in different vision areas, without harming the QoE. For example, we can display the content with highest quality to provide the best viewing experience in central vision, and the contents with lower qualities in other visions to reduce the video size. Since human is less sensitive to the video quality outside central vision area, we will not hurt the QoE.

Specifically, only 33.33% (120/360) of the video, FoV, is horizontally visible to people. Among the FoV, only 1.39% (5/360) of the video can be clearly perceived and 16.67% (60/360) of the video can be seen less clearly. The rest 66.67% of the video is totally invisible. Hence, if we can compose a hierarchical video with mixed qualities, we can significantly reduce the video size.

### C. Video Scheduling

Prefetching is necessary for mobile 360° video streaming. Due to the transmission delay and coding/decoding delay, we cannot instantly fetch a video clip, even for the small central vision. Hence, scheduling the video composition and transmission based on the predicted user attention can help to reduce the time of viewing interrupt.

Figure III presents an illustration example. At time $i$, we are going to display clip $i$ to the user, if we have prefetched clip $i$ based on prediction then everything works well. But if we request clip $i$ only after we get the tracking data and learn the user attention area, then a mismatch arises because of latency. Hence, prefetching video clips is necessary to display video in time and provide a satisfied QoE. Then the prediction of user attention is crucial to guarantee the prefetched video is the wanted. Otherwise, the bandwidth is wasted and QoE is also impaired.

## IV. FoVR DESIGN

In this section, we present our designs of FoVR, a hierarchical 360° video streaming system that leverages the hierarchy of human vision and provides the mixed-quality video to enable a high quality 360° video streaming with satisfied QoE. We first introduce the framework of FoVR and then present the user attention prediction method. Based on the predicted user attention, we present how to compose and schedule the mixed-quality video to achieve satisfied QoE on mobile VRs.

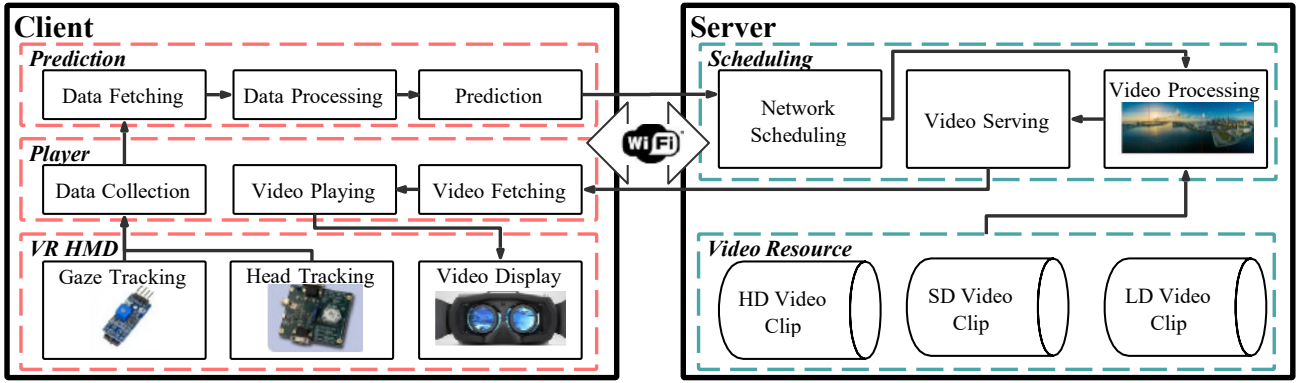

Fig. 4. Overview of FoVR

## A. Overview

Figure 4 presents the framework of FoVR. Our system uses a C/S (client/server) architecture. The client consists of VR HMD, video player and prediction modules. The server consists of scheduling module and the video resources from the providers. A client is connected to the server via the wireless networks.

The whole system can be divided several parts, 1) The VR HMD tracks the user's head and gaze movement information for attention prediction module. 2) After collecting the movement information, the predication module processes the data and performs our SVR-based predication algorithm (Section IV-A2). 3) The predicted results are then transmitted to the server for further video scheduling. With the information of the available bandwidth and user's attention, the server composes the mixed-quality video clips that have the hierarchical definitions according to the hierarchy of human vision. 4) The composed video clips are then scheduled to transmit to the client player that decodes and displays the mixed-quality video to the user.

*1) Data Processing:* We connect VR HMD and get date from sensors via Unity3D [25]. From sensors, we get a ray denoting gaze direction, which includes two 3D vectors representing the start point and the direction of the gaze. We also get a quaternion [26] denoting head orientation, which is a 4D vectors representing rotation. We can calculate unit vectors of head and gaze, then calculate the intersects of head, gaze vectors with the 360° video in the coordinate of 360° videos. Furthermore, the video will be cropped into $m \times n$ tiles. Then we can establish mapping between intersects with tiles in next processing.

*2) Online User Attention Prediction:* In FoVR, we use the Fove VR [22] as the VR HMD. Fove provides IMU based orientation tracking and IR-based position tracking. Besides, it also provides eye tracking with error less than 1°. We leverage the IMU sensors and the eye tracking results to further predict the user attention.

The same as FoV, attention is also a pyramid centring on gaze direction which can be projected as a area of 360° videos, equal in size with the central vision as introduced in Section III-B.

It is impossible to obtain a comprehensive model that satisfies all the conditions and achieves optimal performance in every specific condition. Therefore, in FoVR, both the training and predicating are conducted online to adjust to the constantly changing user movement. We use a moving window that contains the movement data in the most recent $5s$ to fit an attention transformation pattern. Then we leverage the pattern to predict the movement in the following $1s$. The prediction result is a position sequence of projected attention on the video, but we cluster them into a mean position in order to reduce delay of processing. As there are differences between mean and actual attention and errors in prediction, we give a total 5° tolerance to attention. Furthermore, we hierarchically introduce sub-attention and non-attention area around attention which will correspond to different definitions in mixed-quality videos.

We use SVR (Support Vector Regression) [27] with RBF (Radial Basis Function) kernel as the predication model. SVR is promoted from SVM (Support Vector Machine). SVR transforms the sample data into a high-dimensional feature space by nonlinear transformation and constructs a linear decision function in that space. Suppose there is a set $D = (x_1, y_1), (x_2, y_2), ...(x_n, y_n)$, SVR tries to learn a function $f(x) = w \varphi(x) + b$ to make $f(x)$ close to $y$, where $w$ and $b$ are the parameters need to be trained. The objective function can be represented by the following equation.

$$\frac{1}{2} w^2 + \frac{1}{n} \sum_{i=1}^{n} |f(x_i) - y_i|_C \qquad (1)$$

where $|f(x_i) - y_i|_C$ is the loss function. To solve this problem, we can introduce slack variables and Lagrange multipliers to transform the problem to its dual problem, which is a convex quadratic programming problem. By solving the dual problem,

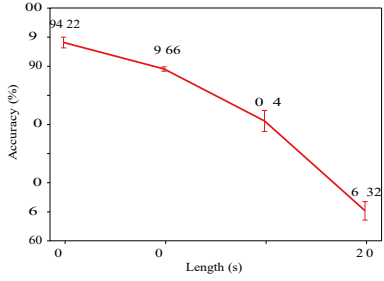

Fig. 5. Prediction accuracy under different prediction lengths.

we get the SVR model:

$$f(x) = w \times \varphi(x) + b = \sum_{i=1}^{n}(a_i^* - a_i)K(x_i, x) + b \quad (2)$$

where $K(x_i, x) = \Phi(x_i) \times \Phi(x)$ is the kernel function. In our current design, we utilize the Gauss RBF function as the kernel function which can be expressed as follows.

$$K(x_i, x) = \exp(-\gamma \|x_i - x\|^2) \quad (3)$$

In the prediction module, we choose $1s$ as time window to predict the user attention due to the following reasons. On the consideration of human cognition, the user generically focuses on the dynamic objects when watching videos [28]. According to the measurement study [29], the dissimilarity between frames will significantly increase after 30 frames apart, which is usually $1s$ in the common video. On the consideration of prediction utility, a too short or too long predicting window will hurt system performance. A short predication window can of course obtain a high predication accuracy due to the temporal correlation of gaze and head movements. But a short window also means we have to continuously compose short video segments which requires high computational capacity on both server and client. On the other hand, a long predication window may have high prediction error rate and lead to the bandwidth waste or even display interrupt because of perfecting the undesired videos, hurting the QoE.

We conduct an experiment to show $1s$ is an appropriate setting to balance the trade-off. We continuously collect the gaze and head movement data for each $5s$ with a sampling rate of 10Hz as training data. Then we test the mean predication accuracy with different prediction window lengths. Figure 5 show the experiment results. As expected, accuracy decreases with the increase of the window length. Using $1s$ can provide an acceptable accuracy for further scheduling to ensure the QoE. Hence, in our current design and implementation, we set prediction window length as $1s$.

### B. Composition of Mixed-quality Videos

FoVR transmits mixed-quality video to the client for bandwidth and computing resource consumption reduction, based on the user attention predication. To reduce the delay of online video processing, we first perform offline video slicing and store the video clips with different definitions in the video resource library. Then we can directly request the video clips with desired definition and perform online integration, which is much faster. Figure IV-B illuminates the processes of offline slicing and online integration of the mixed-quality videos.

We first process the high bitrate videos offline into three types of definitions with H.264 encoding: HD (high-definition), SD (standard-definition) and LD (low-definition). The HD video has a bitrate that equals to the original video. The SD video has a slightly lower bitrate, compared with the original video. The LD video has the lowest bitrate to greatly reduce the video size. In addition, SD and LD videos will have multiple bitrates, which are used to adapt to the dynamic network conditions. With multiple bitrates, FoVR can select and schedule the optimal video composition and transmission strategy according to the current network condition.

We then slice the video into many clips, each clip is $1s$ long, consistent to the predication window length. To slice videos, we need restructure the Group of Pictures (GOP) which make up videos. A GOP consist several keyframes (I), forward-predicted frames (P) and bi-directionally predicted frames (B). P frames make prediction from I and B frames make prediction from both I and P frames. A closed-GOP begins with a I frame and refers to the frames in the same GOP to generate the other frames. But an open-GOP does not necessarily begin with a I frame and may generate the frames based on adjacent GOPs. In FoVR, we desired the sliced clips have the same length with frame I as the first frame and every clip is independent to easily schedule and compose the mix-quality video. Therefore, we restructure the videos into closed GOP form to ensure we get independent clips with the same length.

Finally, each video clip is cropped into $m \times n$ grid, as mentioned in Section IV-A1. We choose $m = 36$ and $n = 36$ in our implementation. Because the most clear central vision for gaze is about 5° as described in Section III-B, besides we introduce a 5° tolerance as discussed in Section IV-A2 to bear the attention prediction errors for a better QoE. So each tile corresponds to 10° of the whole 360° video. Then the attention area with HD is mapping to a tile with 10°, the sub-attention area with SD is mapping to about $3 \times 3$ tiles, the non-attention area with LD is mapping to the rest. These three areas correspond to attention, sub-attention and non-attention, respectively. The online composition of mixed-quality video will be introduced together with the scheduling in the next section because the composition is coupled with the scheduling to provide better performance. In the process of slicing and cropping, we introduce additional storage usage for tiles with multiple definition. Compared with expensive bandwidth resources, storage is a cheap and accessible resource. It is feasible to relieve bandwidth demands using storage.

### C. Scheduling

Based on the current network condition and the user attention, the network scheduling module dynamically select the suitable video clips with different definitions to compose the mixed-quality video with the highest estimated QoE.

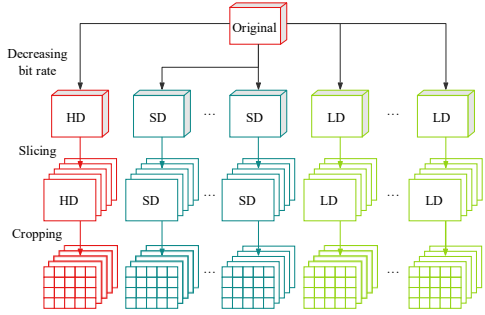

Fig. 6. Offline video processing to get video clips and tiles.

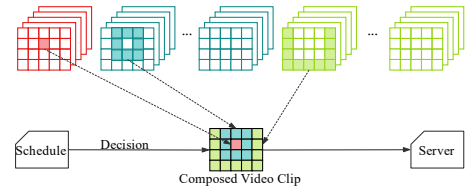

Fig. 7. Real-time video processing to compose a video clip based on scheduling.

*1) The definition of QoE:* We adopt MOS (Mean Opinion Score) [30], the most common QoE metric, as the QoE metric in FoVR. MOS is calculated as the mean of subjective scores in a subjective quality evaluation test as Equation 4 shows.

$$MOS = \frac{\sum_{n=1}^{N} R_n}{N} \quad (4)$$

where $R$ is individual score from subject $n$.

The MOS score is measured by a subjective quality evaluation test which we cannot easily get while streaming. But we can estimate it by the objective parameters, such as bitrate. We use two datasets [31], [32] to explore the relationship between MOS and bitrate. We find that the relationship between bitrate and MOS score meet an exponential function. With curve fitting and normalization, we get an approximate representation of Bitrate-based Video Quality Assessment (BVQA), as shown in Equation 5, which is calculated for each tile, where $x$ denotes the bitrate of tiles in unit of kbps.

$$BVQA = \begin{cases} 1 - e^{-0.648 \times 10^{-3} x} & \text{for LD areas;} \\ 1 - e^{-0.324 \times 10^{-3} x} & \text{for SD areas;} \\ 1 - e^{-0.081 \times 10^{-3} x} & \text{for HD areas.} \end{cases} \quad (5)$$

Then we can define our QoE metric for a video clip as follows.

$$QoE = \sum_{i,j}^{i=N_x, j=N_y} BVQA_{ij} \times Weight_{ij} \quad (6)$$

where $i, j$ is the index of tiles, $N_x, N_y$ is the number of tiles in the vertical and horizontal direction, respectively. We define the empirical weight for HD, SD and LD areas as 0.5, 0.3, 0.2 respectively. Then the concrete weight of each tile will depend on the number of tiles in each area.

*2) Scheduling Algorithm:* We formulate the scheduling problem in FoVR as a Knapsack problem. Specifically, we have a set $T = \{X_1, X_2, ..., X_n\}$ of $n$ video tiles, and each $X_i$ is a set $\{D_1, D_2, ..., D_m\}$ of $m$ kinds of definitions with weight $w_{ij}$ referring to the bandwidth demand of video tile $X_i$ with definition $D_j$. $x_i$ denotes if tile $X_i$ is selected, if yes then $x_i = 1$ otherwise $x_i = 0$. $d_j$ denotes if definition $D_j$ of tile $X_i$ is selected, calculated the same as $x_i$. $v_{ij}$ refers to the calculated BVQA value of tile $X_i$ with definition $D_j$ which can be calculated by prediction results. We assume the whole available bandwidth as the weight capacity $W$. Then we formulate the optimization problem as follows.

$$\begin{aligned} \text{maximize} \quad & QoE = \sum_{i=1}^{n} \sum_{j=1}^{m} v_i x_i, \\ \text{subject to} \quad & \sum_{i=1}^{n} \sum_{j=1}^{m} w_{ij} d_j \in \{0, W\}, \\ & \text{and} \sum_{k=1}^{m} d_k = 1 \end{aligned} \quad (7)$$

Note that, in the formulated problem, we always have $x_i = 1$ and $\sum_{k=1}^{m} d_k = 1$ to ensure the integrity of video in our scheme.

To solve the optimization problem, we propose a greedy algorithm. We first select the highest definition for all three areas, regardless of the available bandwidth. Then we lower the definition of the LD areas, which can be determined by the user attention prediction module. If the available bandwidth is still insufficient, we then lower the the definition of SD areas. We recursively execute the process until we can fit the video in the available bandwidth. During the process, we do not reduce the definition of HD area, because users focus on this area and reducing the definitions of this area will harm QoE. Instead, we adjust the bitrates of SD and HD areas to adapt to the limited bandwidth.

*3) Composing Video clips:* After deciding the definition for each tile, FoVR selects the video tiles from corresponding video clips and compose the final video clip using these tiles. Figure IV-B shows the processing workflow. Our minimal processing unit is $1s$ tiles sliced in offline process.

## V. IMPLEMENTATION

We implement FoVR on the off-the-shelf devices. Our client consists of a Fove VR HMD [22] and a Windows host. The Fove tracks the user's head orientation through the IMU and tracks the user's head position based on the IR sensor. The VR HMD uses two IR-based eye tracking sensors to provide eye tracking service with error less than 1°. Note that we can eliminate the Windows host because our design does not rely on any specific function of the host. We implement the modules on the Windows host instead of the VR HMD only because the used VR HMD does not support custom development. It will be easy to transplant our implementation to the VR HMD that supports custom development. A PC running Ubuntu acts as the server of FoVR. On the server, we build a DASH server to serve video clips. We implement the scheduling algorithms using Python, and process videos under the help of OpenCV. The client and server use 802.11ac wireless network for communication.

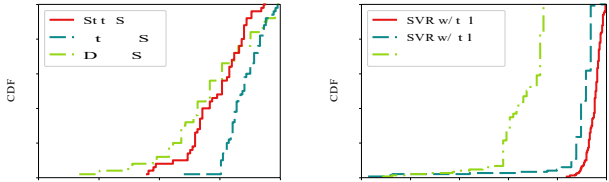
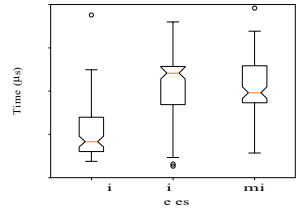
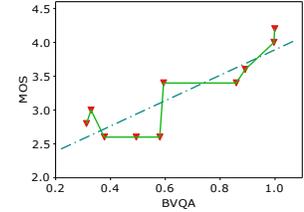

Fig. 8. Prediction accuracy in different scenes.  Fig. 9. Prediction accuracy w/ different models.  Fig. 10. Time delay of each prediction.  Fig. 11. Relation between BVQA and MOS

On the client, we develop a 360° video player based on Unity [25] to play 360° video streaming. Specifically, we create a sphere model in Unity3D, and use inner surface of the sphere model as screen to display 360 videos. We place the camera, i.e., the user's FoV, in center of the sphere, and connect the camera to the VR HMD via the VR SDK to synchronize the motion of VR HMD with the motion of the camera. In terms of codec, we use VLC video library [33] replacing the Unity3D's own Movie Texture, to gain audio support for video and improve the performance and efficiency of video playing.

Since the video player is the only module directly connected to the HMD, we program the player to help collect the head and gaze movement data provided by VR sensors, and transmits the data to the prediction module. We implement the attention prediction module on the client host in Python. This module fetches data from the player, performs prediction and then transmits the results to the server.

## VI. EVALUATION

In this section, we evaluate the performance of FoVR in various scenarios. We adopts three types of videos. The first type is Static Scene in which the scene is static and only the camera moves. The second type is Motion Scene in which no more than three objects moves in the scene. The third one is Dynamic Scene in which everything is moving. We study the performance of each component of FoVR and evaluate the overall improvement of FoVR.

### A. Attention Prediction

To investigate the performance of our attention prediction algorithm, we ask five volunteers to watch videos and record their head and gaze movements. Then we use the collected data as a time series to predict users' attention. We divide each set of data into a number of consecutive time series with a length of $6s$. For each time series, we use the first $5s$ data to predict the user's attention in the last $1s$.

*1) Accuracy:* Figure 8 presents the CDF of prediction accuracy in three scenarios. The average accuracy is 94.71%, 97.01% and 94.18% for static, motion and dynamic scenarios, respectively. We can see that we get the best prediction on Motion Scene. It's because there are several main objects in the scene, users' movement will trend to follow these objects, which result to a more regular movement so that it's more simple to predict. As for the Static Scene and Dynamic Scene, users will watch according to their own interests, so the movement pattern can be less regular for prediction. With $5°$ error tolerance, even in the unfavorable scenarios, FoVR can achieve a prediction accuracy of 90% for more than 95% cases.

We also compare the prediction model in FoVR with other some recent models that can be used for prediction. From Figure 9, we can find that FoVR achieves 95.30% accuracy on average, which is about 30% high than linear regression that achieves 65.83% accuracy. Even for the SVR without tolerance, the achieved accuracy is 89.66% on average. But it generates wrong predictions on about 5% cases. Our prediction achieve more than 80% accuracy even in the worst case.

*2) Delay:* We also study the computing delay of prediction. Figure 10 shows the average time required for attention prediction is $65.89\mu s$. The difference among three scenarios and the jitters in each scenario are due to random errors by CPU. The computation overhead in the $\mu s$ level is quite small, compared with the $1s$ length of video clip. Hence, it has no influence on streaming delay. The experiment results demonstrate our predication algorithm can provide accurate attention information in a short time for further scheduling in various video scenarios.

### B. Scheduling

*1) Subjective Experiment:* We perform a subjective experiment to validate if our metric BVQA correlates with users' subjective opinion. We play 10 videos with different definitions in the same random order to 5 volunteers can record their MOS on these videos to compare with BVQA values. Figure 11 shows the result, the Pearson correlation coefficient of these two metric are 0.8637, which indicates our metric correlates with users' subjective opinion.

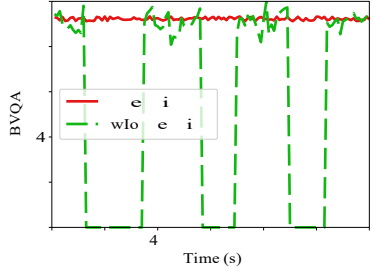
(a) 10Mbps bandwidth.
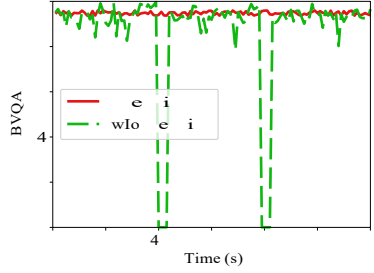
(b) 50Mbps bandwidth.
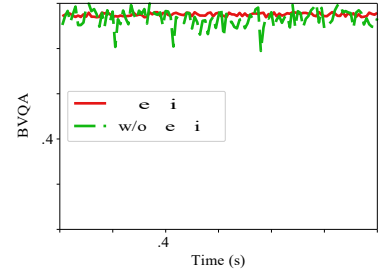
(c) 100Mbps bandwidth.

Fig. 12. QoE evaluation by BVQA under different bandwidths.

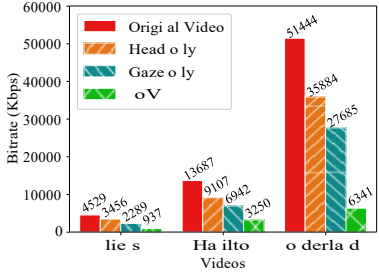
Fig. 13. Bitrate comparison with different composition schemes.

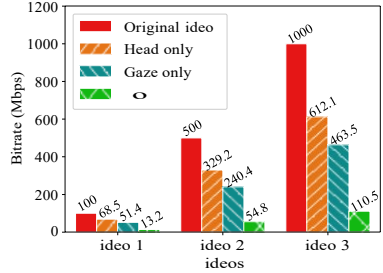
Fig. 14. Bitrate comparison with different composition schemes using simulation videos.

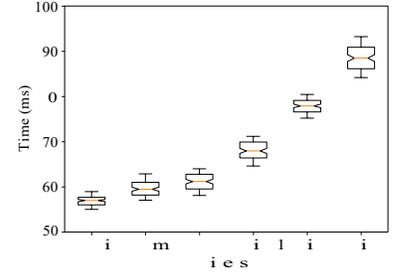
Fig. 15. Compression ratio and delay of video processing.

*2) QoE Evaluation:* We evaluate our scheduling algorithm using BVQA defined in Section IV-C. We limit the network bandwidth to 10Mbps, 50Mbps and 100Mbps to simulate different network conditions. We played videos for 2min and calculate the BVQA in different settings. For comparison, we implement a method without scheduling that arbitrarily chooses tiles to compose the video clips.

Figure 12 shows the BVQA values obtained in different bandwidths. Under the scenario with 10Mbps bandwidth, the average BVQA is 0.992 and 0.538 for methods with and without scheduling, respectively. Figure 12a shows the BVQA of non-scheduling approach experiences stalling because the composed clip is too large to be transmitted in time under the limited bandwidth. In the scenario with 50Mbps bandwidth, our scheduling method achieves 0.947 BVQA and non-scheduling approach gets 0.871 BVQA on average. The non-scheduling approach meets stalling as well but with less times, because a higher bandwidth has higher transmitting capacity. For 100Mbps bandwidth, the BVQA is 0.950 and 0.934 for methods with and without scheduling, respectively. There is no stalling for both approaches due to the high bandwidth. However, the BVQA with scheduling is more stable than non-scheduling. The standard deviations are 0.006, 0.007, 0.007 with scheduling and 0.448, 0.236, 0.038 without scheduling, respectively under three different scenarios. The evaluation results demonstrate our scheduling algorithm can ensure a high and stable BVQA, indicating our scheduling method can provide users with a favourable QoE.

### C. Video composition

We selected three videos from our video dataset with different bitrates for composition evaluation. The metadata of the selected videos are shown in Table I. We process the bitrates of SD clips and LD clips to achieve more than 0.9 BVQA value for each videos.

TABLE I
METADATA OF VIDEOS USED IN EXPERIMENT

|  | Resolution | Size(MB) | Bitrate(Kbps) | Dynamics |
|---|---|---|---|---|
| Aliens | 1920x960 | 156 | 4529 | Static |
| Hamilton | 3840x2048 | 767 | 13687 | Little Motion |
| Wonderland 3 | 3840x2160 | 802 | 51444 | Dynamic |

*1) Compression ratio:* We perform video processing as mentioned in Section IV. We compare our scheme with state-of-the-art head-only scheme (used in tile-based streaming) and gaze-only scheme. For head-only scheme, we compose the area corresponding to head direction with HD definition and the rest areas with SD definition. Equivalently, we compose the area corresponding to gaze direction with HD definition and the rest areas with SD definition for gaze-only scheme. We perform each experiments for 10 times and then calculate the mean value of bitrate.

Figure 13 shows that our scheme achieve the best compression ratio with mean value 81.08% while the other two schemes only get 29.13% and 48.31% on average. The reason that we can achieve high compression ratio is that the proportion of HD and SD tiles in our scheme is much lower than LD tiles, the bitrate of the composed videos are very close to the whole bitrate of LD tiles which is quite low. Meanwhile the bitrate of videos from current streaming providers is still too low for VR-ready 840Mbps bitrate, so we generate some high bitrate videos to verify if our scheme can still achieve the same high compression ratio. We re-encode the Wonderland video to simulate high-bitrate videos.

Figure 14 shows that the compression ratio is 34.82%, 51.40% and 88.26%, respectively using head-only scheme, gaze-only scheme and our hierarchical scheme, which is consistent with the real-world evaluation.

*2) Delay:* We also calculate the delay of composing 1s video clip. We evaluate both real-world videos and simulation videos. Figure 15 shows the experiment results. The average processing time is 56.96ms, 59.67ms, 61.14ms and 68.09ms, 77.84ms, 88.62ms for real-world and simulation videos, respectively. The bitrate of original videos increases along with the delay. Because a video with high bitrate contains more detailed information so that this video needs more time to compute. But the longest delay is no more than 90ms for processing video with the 1Gbps bitrate. Counting in the prediction delay and composition delay, more than 900ms are left for starting prefetching. As for communication delay, this delay is quite low (usually dozens of ms) due to the effort of CDN. Even if the communication delay fluctuates up to hundreds of ms there is still enough time left for prefetching. Therefore, the delay of video processing in FoVR is tolerable.

The results show that our compression scheme achieves a high compression ratio with a short time delay, which ensures that our streaming can fit current network condition.

## VII. CONCLUSION

In this work, we study the problem about how to enable high-quality 360° video streaming on untethered mobile devices without harming the QoE. The key challenge is the mismatch between the limited bandwidth capacity of existing wireless networks and the high bandwidth demands of high-quality 360° video streaming. We propose and implement FoVR, a practical 360° video streaming system that leverages the hierarchy of human vision to lower the video qualities of the unneeded fields. By providing the mixed-quality video, FoVR significantly reduces the video size and the bandwidth demand as well. FoVR also integrates an online attention prediction algorithm that leverages the head and gaze movements to predict the user attention field where the high quality video should be displayed. Based on the predicted user attention, FoVR further schedules the the composition and transmission of videos to reduce the service delay and improve the QoE. The evaluation results show that FoVR can reduce the bandwidth demands by more than 88.9% and 76.2%, compared with the original VR video and the state-of-the-art method respectively. Thanks to the significant reduction of video size and the efficient online scheduling, FoVR is able to play the video smoothly and adapt to different network conditions with the mean BVQA of 0.95.

## ACKNOWLEDGEMENTS

This work is support in part by National Key R&D Program of China No. 2017YFB1003000, National Natural Science Foundation of China No. 61772306, No. 61672372, No. 61672240. We thank all the anonymous reviewers for their valuable comments and helpful suggestions.

## REFERENCES

[1] "Oculus rift," https://www.oculus.com/rift/.
[2] "Htc vive," https://www.vive.com/cn/.
[3] "Google daydream," https://vr.google.com/daydream.
[4] Y. He, J. Guo, and X. Zheng, "From surveillance to digital twin: Challenges and recent advances of signal processing for industrial internet of things," *IEEE Signal Processing Magazine*, vol. 35, no. 5, pp. 120–129, 2018.
[5] "Global virtual reality (vr) market set for rapid growth, to reach around usd 26.89 billion by 2022," https://www.zionmarketresearch.com/news/-virtual-reality-market.
[6] "2016 global phenomena: Latin america & north america," https://www.sandvine.com/blog/2016/06/global-internet-phenomena-report-2016-latin-america-north-america.
[7] "Vr big data report 2016," https://www.huawei.com/˜/media/-CORPORATE/PDF/ilab/19-en.
[8] Z. Zhou, C. Wu, Z. Yang, and Y. Liu, "Sensorless sensing with wifi," *Tsinghua Science and Technology*, vol. 20, no. 1, pp. 1–6, Feb 2015.
[9] "Ieee std 802.11ac-2013," http://standards.ieee.org/findstds/standard/802.-11ac-2013.html.
[10] "Tpcast," https://www.tpcastvr.com/.
[11] "Display link xr," http://www.displaylink.com/vr.
[12] O. Abari, D. Bharadia, A. Duffield, and D. Katabi, "Enabling high-quality untethered virtual reality." in *Proceedings of USENIX NSDI*, 2017.
[13] F. Qian, L. Ji, B. Han, and V. Gopalakrishnan, "Optimizing 360 video delivery over cellular networks," in *Proceedings of ACM ATC@MobiCom*, 2016.
[14] M. Hosseini and V. Swaminathan, "Adaptive 360 vr video streaming: Divide and conquer," in *Proceedings of IEEE ISM*, 2016.
[15] X. Xie and X. Zhang, "Poi360: Panoramic mobile video telephony over lte cellular networks," in *Proceedings of ACM CoNEXT*, 2017.
[16] J. He, M. A. Qureshi, L. Qiu, J. Li, F. Li, and L. Han, "Rubiks: Practical 360-degree streaming for smartphones," in *Proceedings of ACM MobiSys*, 2018.
[17] K. Boos, D. Chu, and E. Cuervo, "Flashback: Immersive virtual reality on mobile devices via rendering memoization," in *Proceedings of ACM MobiSys*, 2016.
[18] Z. Lai, Y. C. Hu, Y. Cui, L. Sun, and N. Dai, "Furion: Engineering high-quality immersive virtual reality on today's mobile devices," in *Proceedings of ACM MobiCom*, 2017.
[19] L. Liu, R. Zhong, W. Zhang, Y. Liu, J. Zhang, L. Zhang, and M. Gruteser, "Cutting the cord: Designing a high-quality untethered vr system with low latency remote rendering," in *Proceedings of ACM MobiSys*, 2018.
[20] B. Guenter, M. Finch, S. Drucker, D. Tan, and J. Snyder, "Foveated 3d graphics," *ACM TOG*, vol. 31, no. 6, 2012.
[21] "Tobii," https://www.tobii.com/.
[22] "Fove vr," https://www.getfove.com/.
[23] "Binocular vision," https://en.wikipedia.org/wiki/Binocular_vision.
[24] M. S. DoD, "Department of defense design criteria standard: Human engineering (mil-std-1472g)," *Department of Defense, Washington*, 2012.
[25] "Unity3d," https://unity3d.com/.
[26] K. Shoemake, "Animating rotation with quaternion curves," in *ACM SIGGRAPH computer graphics*, vol. 19, no. 3. ACM, 1985, pp. 245–254.
[27] A. J. Smola and B. Scholkopf, "A tutorial on support vector regression," *Statistics and computing*, vol. 14, no. 3, 2004.
[28] P. K. Mital, T. J. Smith, R. L. Hill, and J. M. Henderson, "Clustering of gaze during dynamic scene viewing is predicted by motion," *Cognitive Computation*, vol. 3, no. 1, pp. 5–24, 2011.
[29] C.-L. Huang and B.-Y. Liao, "A robust scene-change detection method for video segmentation," *IEEE transactions on circuits and systems for video technology*, vol. 11, no. 12, pp. 1281–1288, 2001.
[30] "Vocabulary for performance and quality of service," https://www.itu.int-/rec/T-REC-P.10.
[31] Y. Zhu, L. Song, R. Xie, and W. Zhang, "Sjtu 4k video subjective quality dataset for content adaptive bit rate estimation without encoding," in *Proceedings of IEEE BMSB*, 2016.
[32] M. Cheon and J.-S. Lee, "Subjective and objective quality assessment of compressed 4k uhd videos for immersive experience," *IEEE TCSVT*, vol. 28, 2018.
[33] "Vlc media player," https://videolan.org/vlc/.